\documentclass[conference]{IEEEtran}
\IEEEoverridecommandlockouts

\usepackage{cite}
\usepackage{mwe}
\usepackage{subfig}
\usepackage{setspace}
\usepackage{algorithm}
\usepackage{algorithmic}
\usepackage{float}%
\usepackage{commath}
\usepackage{ifpdf}
\usepackage{cuted}
\usepackage{lipsum}
\usepackage{diagbox}
\usepackage{multirow}
\usepackage{romannum}
\usepackage{etoolbox}
\usepackage{mathtools}
\usepackage{amsfonts, amsmath, amsthm, amssymb}
\usepackage{bbm}
\usepackage{graphicx, wrapfig, color}
\usepackage{hyperref}
\usepackage[normalem]{ulem}

\usepackage{mathtools}
\mathtoolsset{showonlyrefs}

\usepackage{comment}

\newtheorem{problem}{Problem}

\newcommand{\F}{\mathbb{F}}
\newcommand{\R}{\mathbb{R}}

\newcommand{\B}{\mathbb{B}}
\newcommand{\E}{\mathbb{E}}
\newcommand{\Prob}{\mathbb{P}}
\newcommand{\1}{\mathbbm{1}}
\newcommand{\C}{\mathbb{C}}
\newcommand{\Ent}{\mathbb{H}}

\allowdisplaybreaks

\newtheorem{theorem}{Theorem}
\newtheorem{proposition}[theorem]{Proposition}

\allowdisplaybreaks

\usepackage{xcolor}

\begin{document}

\bstctlcite{IEEEexample:BSTcontrol}


\title{Interpreting Training Aspects of Deep-Learned Error-Correcting Codes
}

\author{\IEEEauthorblockN{N. Devroye,  A. Mulgund,  R. Shekhar,   Gy. Tur\'an$^*$, M. \v Zefran, and Y. Zhou}
\IEEEauthorblockA{University of Illinois Chicago (UIC), Chicago, IL, USA}
\IEEEauthorblockA{$^*$ UIC and MTA-SZTE Research Group on Artificial Intelligence, ELRN, Szeged, Hungary}
\IEEEauthorblockA{\{devroye,  mulgund2,  rshekh3, gyt,  yzhou238, mzefran\}@uic.edu}
\thanks{This work was supported by NSF under awards 1934915 and 1900911. Computing resources for the experiments were provided in part by the NSF award 1828265 (COMPaaS DLV).
The authors are in alphabetic order.}}


\maketitle

\begin{abstract} 
As new deep-learned error-correcting codes continue to be introduced, it is important to develop tools to interpret the designed codes and understand the training process. Prior work focusing on the deep-learned TurboAE has both interpreted the learned encoders post-hoc by mapping these onto nearby ``interpretable'' encoders, and experimentally evaluated the performance of these interpretable encoders with various decoders. Here we look at developing tools for interpreting the training process for deep-learned error-correcting codes, focusing on:  1) using the Goldreich-Levin algorithm to quickly interpret the learned encoder; 2) using Fourier coefficients as a tool for understanding the training dynamics and the loss landscape; 3) reformulating the training loss, the binary cross entropy, by relating it to encoder and decoder parameters, and the bit error rate (BER); 4) using these insights to formulate and study a new training procedure.  All tools are demonstrated on TurboAE, but are applicable to other deep-learned forward error correcting codes (without feedback).
\end{abstract}

\section{Introduction}



Coding theory aims to develop optimal encoder-decoder pairs for various channels and optimization criteria. This has traditionally been done more or less ``by hand'' using theoretical insights and mathematical and algorithmic constructions. Recently, however,  this has been attempted with a new twist:
 use machine learning / deep-learning to learn the encoding and/or decoding functions directly~\cite{deep-phy, deep-low, oshea-2016, deep-model-independent,deep-source, deepIC, deepcode, deepjoint, turboAE, chahine2022inventing, gunduz:allyouneed}.
 We refer to such codes as deep-learned error-correcting codes (DL-ECC). The approach has been successful, particularly for channels with feedback~\cite{deepcode, gunduz:allyouneed, chahine2022inventing}, but also for point-to-point Additive White Gaussian Noise (AWGN) channels, one of the benchmarks for practical code performance~\cite{turboAE, Makkuva2021KOCI}. 
 
Deep learning provides computational tools for viewing the task as an optimization problem and solving it efficiently by training a neural network. The rapidly evolving toolkit of deep learning offers many possible
architectures and allows (approximately) optimal codes to be found by using a training procedure.
For example, for channels with feedback, \cite{deepcode} uses Recurrent Neural Networks, while~\cite{gunduz:allyouneed} uses a
transformer-based architecture. For point-to-point channels, \cite{turboAE} mimics the architecture of Turbo-codes, replacing convolutional codes with Convolutional Neural Networks (CNN), while~\cite{Makkuva2021KOCI} uses non-linear learned components in a  Reed-Mueller-like construction.
Using deep learning and directly optimizing over codes raises important new questions: 

1) Can one understand deep learning training procedures in terms of a search process in the space of codes?
{\it We suggest looking at the Fourier expansion of a learned encoder as a way to both understand the final code, and tracking this, as a way to understand the training dynamics. We propose an efficient way to find the dominant Fourier coefficients using the Goldreich-Levin algorithm~\cite{o2014analysis}.} 

2) If we minimize a loss function over a set of codes (those expressible by a given architecture) then what can we say about the loss landscape~\cite{li2018visualizing}? {\it We show that for certain architectures and loss functions, parity functions appear to be minimizers.} 

3) Loss functions are chosen to facilitate the training process (e.g. binary cross-entropy, BCE), but how do they relate to the performance of a code (e.g. bit error rate, BER)? {\it We show tight bounds connecting BCE and BER.} 

4) If optimization is viewed as working over the two-dimensional (encoder, decoder) space, what can be said about the structure of this space?  For example, several training procedures~\cite{turboAE, deepcode} alternate training the encoder and decoder -- is this fundamentally needed, or just a
practical
way to improve convergence? Is there a way to decompose the loss function into an encoder-only and a decoder-only component and exploit this? {\it We show that there is such a decomposition and suggest an approach that exploits this.} 


We attempt to study these issues through a mixture of theoretical and experimental insights, focusing on the specific DL-ECC termed TurboAE~\cite{turboAE}, which has been one of the few DL-ECCs for which interpretability studies have been initiated~\cite{US-ISIT-2022, US-Allerton-2022}. Some of the questions we seek to answer here were inspired by passing remarks in~\cite{US-ISIT-2022} and \cite{US-Allerton-2022}; we expand upon these in the next sections. Our experimental observations raise several algorithmic and theoretical questions.
This research direction aims to connect coding theory with machine learning interpretability and deep-learning theory.

\section{TurboAE: basics and past interpretations}

\begin{figure}
    \centering
    \includegraphics[width=0.45\textwidth]{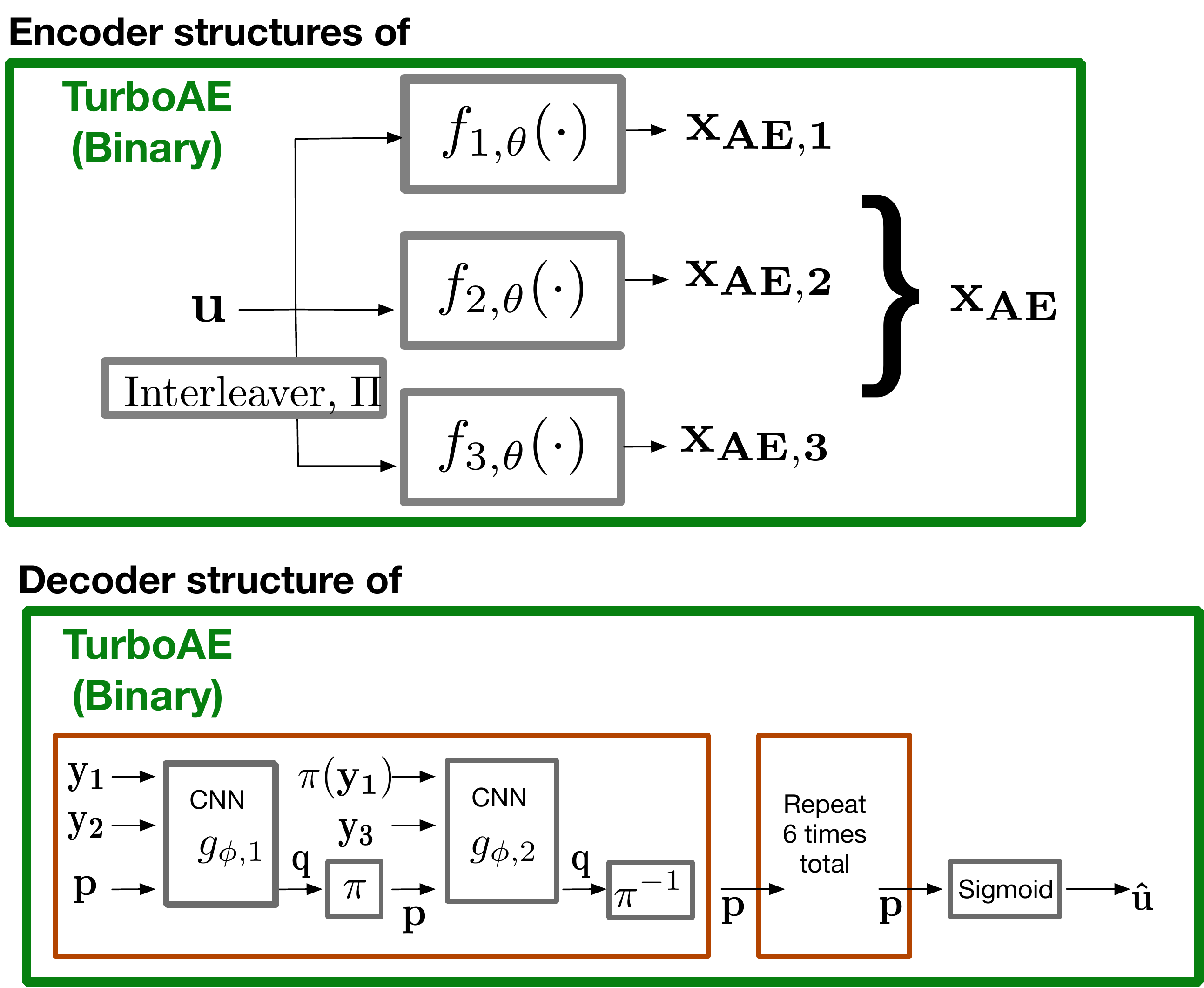}
    \caption{\small Above is the rate $R=\frac{1}{3}$ 
    (${\bf u} \in \mathbb{F}_2^{100}, {\bf x}_{AE,j}\in \{\pm 1\}^{100}$) 
    TurboAE-binary encoder structure.  Functions $f_{j,\theta}(\cdot)$ are the constituent codes implemented as CNNs. 
 Below is the high level decoder structure of TurboAE remade from \cite{turboAE}. The parameters of the CNNs $g_{\phi,1}, g_{\phi,2}$ are trained. The noisy channel outputs of the three encoded streams ${\bf x_1, x_2, x_3}$ are given by ${\bf y_1, y_2, y_3}$. The interleaver is $\pi$ (and its inverse $\pi^{-1}$). The decoder produces probabilities that each input bit is $0$ or $1$, denoted by $\hat{{\bf u}}$.}
    \label{fig:encdec}
    \vspace{-5mm}
\end{figure}

The architecture of the DL-ECC termed ``TurboAE'' encoder network~\cite{turboAE} is based upon a classical rate $\frac{1}{3}$ Turbo code, with the three ``constituent codes'' 
replaced by CNN blocks $f_{b,\theta}(\cdot)$, $b\in \{1,2,3\}$ as in Fig. \ref{fig:encdec} (adapted from~\cite{US-ISIT-2022}). 
Similarly, the TurboAE decoder architecture replaces the iterations of the BCJR decoder by CNNs $g_{\phi, 1}, g_{\phi, 2}$ as in Fig. \ref{fig:encdec} (adapted from~\cite{turboAE}), where ${\bf y_b} \in \mathbb{R}^{100}$ (we use bold font for vectors) and ${\bf y_b} = {\bf x_{\text{AE}, b}} + {\bf z_b}$, for ${\bf z_b}$ i.i.d. Gaussian noise of mean zero and variance 1, for each stream $b\in\{1,2,3\}$.  The network is trained in an end-to-end fashion to obtain the network parameters of the encoder and decoder CNNs jointly.

The input to the network is a sequence ${\bf u}$ of $100$ bits, and the output of each block $b \in \{1,2,3\}$ is a sequence 
${\bf x}_{AE,b} \in \{\pm 1\}^{100}$.  The network has two versions, TurboAE-cont (with real-valued encoder outputs, essentially performing coding and modulation tasks jointly) and TurboAE-binary (with Boolean encoder outputs which are then modulated for transmission over an AWGN channel).
The power control modules are omitted, as is the treatment of the boundary first 2 and last 2 bits, discussed in \cite{US-Allerton-2022} (not needed here).


In \cite{US-ISIT-2022} ``interpretation'' of TurboAE-binary was attempted through both exact (non-linear) and approximate (linear, or parity) approximations of the encoding functions $f_{i,\theta}(\cdot)$; these are provided in Tables \ref{table:bestLinearApprox} and \ref{table:exact} in the Appendix. From these, we see that TurboAE-binary's encoders are non-recursive, non-systematic, and 
$f_{1,\theta}, f_{3,\theta}$ are  non-linear and $f_{2,\theta}$ is a linear function of the 5 inputs at times ${j-2}, \cdots {j+2}$.



In \cite{US-ISIT-2022}, besides finding the exact and best linear approximations to the  encoder functions, several other ``interpretation'' tools were suggested but not deeply explored. Among these is 1) the use of the Fourier representation of Boolean and pseudo-Boolean functions to better understand the training dynamics, and 2) 
the suggestion of using the Goldreich-Levin algorithm to find the largest Fourier coefficient(s) as an approximation algorithm for the encoder. We expand on these here, and also look more deeply into the training dynamics, offering an alternative training to that presented in the original TurboAE \cite{turboAE}. All code will be posted on github if the paper is accepted.


\section{Interpretations through the Fourier lens}
We will later investigate the tracking of Fourier coefficients (FC) of the encoder as a tool for understanding both the training dynamics and the loss landscape. We first discuss an approach to estimate the dominant  FC of the learned encoding functions, which may have a large number of inputs. 

Changing to the domain $x_i \in \{\pm 1\}$, and letting $\chi_S = \prod_{i \in S} x_i$, each Boolean ($\mathbb{F}_2^n \rightarrow \mathbb{F}_2$) and pseudo-Boolean ($\mathbb{F}_2^n \rightarrow \mathbb{R}$) function has a unique \emph{Fourier representation} ~\cite{o2014analysis}  \[ f(x) = \sum_{S \subseteq \{1,2,\cdots, n\}} \hat{f}(S) \chi_S,\] 
where $\hat{f}(S)$ is termed  the FC for set $S$, and $|\hat{f}(S)|^2$ represents the Fourier weight.


The Goldreich-Levin algorithm~(GL)~\cite{goldreich1989hard} seeks to output a list of sets $S$ 
for which $|\hat{f}(S)|$ is larger than a pre-specified threshold $\gamma$.  It requires ``query access'', which in our context simply means evaluating the neural network on an input. 

The  theoretical underpinnings are presented in Theorem \ref{th:gl}~\cite{o2014analysis}, and our implementation is detailed in the Appendix.
 We explore the applicability of this algorithm in estimating the largest FCs of the encoder of a deep-learned error-correcting code (first proposed for this purpose in \cite{US-ISIT-2022}), an alternative method to that used in \cite{US-ISIT-2022} for finding the best parity-approximation.  
  The algorithm
  has been used in theoretical domains such as cryptography \cite{haastad1999pseudorandom}, learning theory \cite{kushilevitz1991learning}, and coding theory \cite{akavia2003proving, abdouli_goldreich-levin_2012} (as a randomized list decoder for Reed-Mueller RM(1,m) codes), but practical implementations and experiments appear limited. We explore some practical aspects of this algorithm.

 \begin{theorem}{\cite{o2014analysis}} \label{th:gl}
Given query access to $f:\{\pm 1\}^n \rightarrow \{\pm 1\}$, given $\gamma, \delta > 0$, there is a $poly(n, \frac{1}{\gamma}\log \frac{1}{\delta})$-time algorithm that outputs a list $L = \{S_1,\cdots,S_m\}$ such that (1) if $\lvert \hat{f}(S) \rvert \geq
\gamma$, then $S\in L$, and (2) if $S\in L$, then $\lvert \hat{f}(S) \rvert \geq \frac{\gamma}{2}$ holds with probability $1-\delta$.
\end{theorem}

The theorem requires $\gamma$ to be specified in advance, which holds in several applications. In general, $\gamma$ can be exponentially small (e.g., for bent functions~\cite{rothaus_bent_1976} where every coefficient is $2^{-n/2}$ for $n$ inputs), but it is polynomial, e.g., for functions with small $|L_1|$-norm. If $\gamma$ is too high then nothing is returned and if it is too low then the running time increases and more coefficients are returned than desired.
We are not aware of any work which calculates explicit constants for the number of queries needed in Theorem~\ref{th:gl}. 

We apply GL to TurboAE-binary, exploring its feasibility and the number of queries needed.
Previous work~\cite{US-ISIT-2022, US-Allerton-2022} is based on the CNN architecture which suggests that the functions in question depend on 9 (TurboAE-cont) or 5 (TurboAE-binary) variables. Here we do not use such a priori knowledge.

Given each TurboAE-binary constituent code $f_{i,\theta}:\{\pm 1\}^{100} \rightarrow \{\pm 1\}^{100}$, we randomly select one output bit. We implemented a heuristic procedure for determining $\gamma$ for each block and computing the minimum number of queries for each block to get the correct result. Details are given in the Appendix. How to best pick $\gamma$ and a minimal number of queries in a principled way is an interesting open question.

\begin{figure}
\centerline{\includegraphics[width=0.97 \columnwidth]{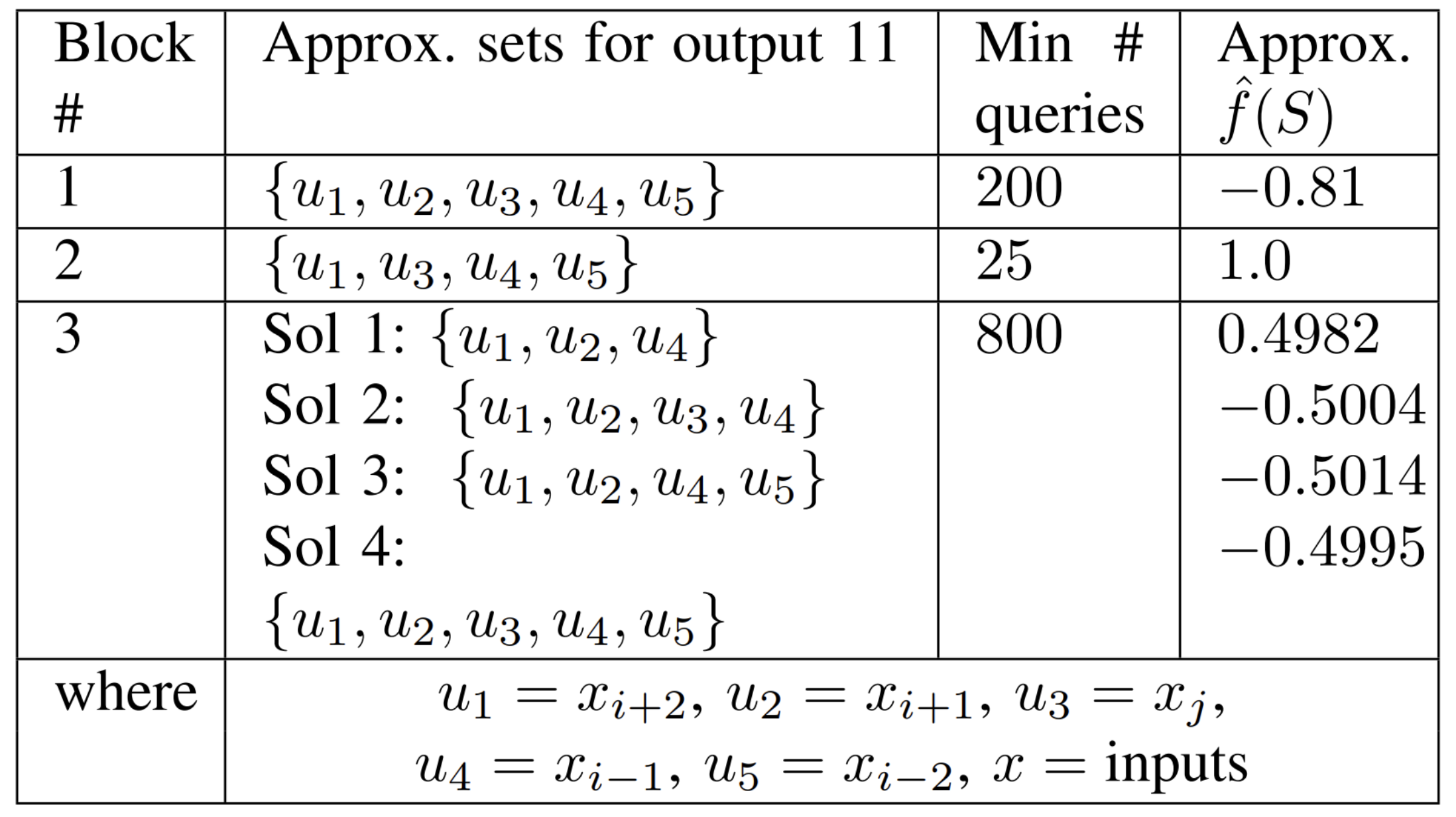}}
   \caption{ \small Goldreich-Levin approximation for TurboAE-binary.}
\label{table:gltable}
\end{figure}

Table in Fig.~\ref{table:gltable}  shows the experimental results. The number of queries refers to the number of function evaluations for estimating a single expectation.

\section{Training dynamics: evolution of Fourier coefficients (FC)}
\label{sec:losslandscape}




Continuing the theme of analyzing Boolean functions in Fourier space, we explore Fourier representation as a tool for understanding the training dynamics and the loss landscape.

\subsection{Dominance and stability of a few Fourier coefficients} \label{section:fc_dominance}
\begin{figure}
\centering
\subfloat[{\small Largest FCs of Block 1 with energy $\geq 95\%$ over 4 independent training sessions. Always a few FCs dominate the Fourier space.}]{\includegraphics[width=0.97 \columnwidth]{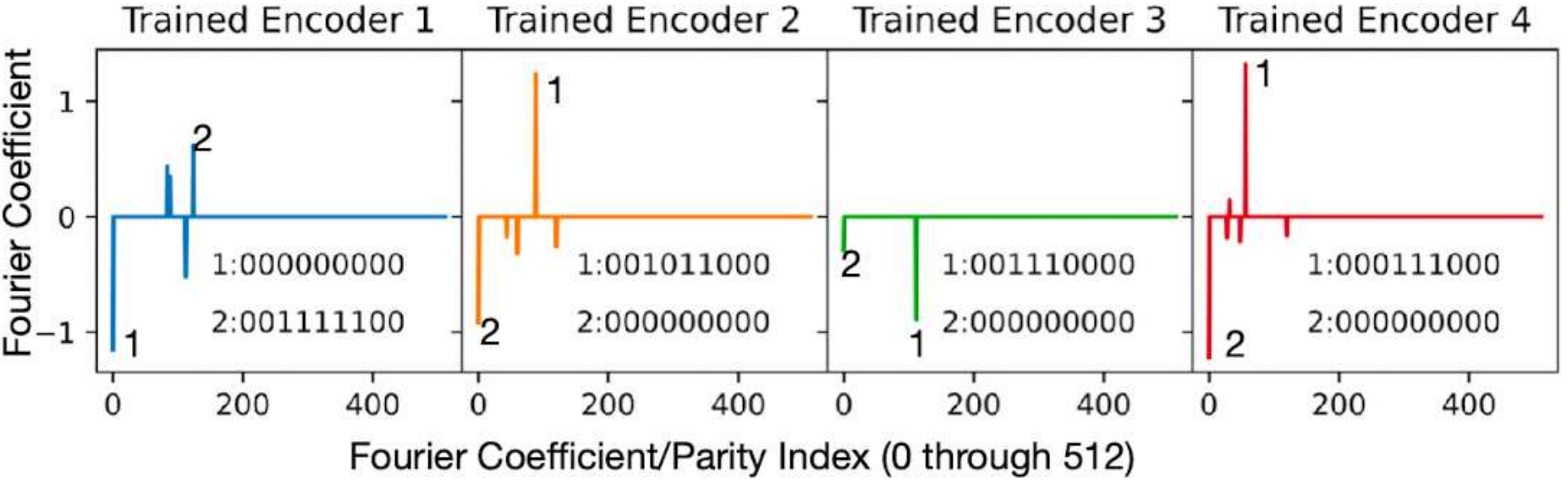}}\\
\subfloat[{\small{FCs of block 3 over different epochs of a training session. Initially, 1-bit parities always dominate, but later higher degree parities emerge.}}]{\includegraphics[width=0.97 \columnwidth]{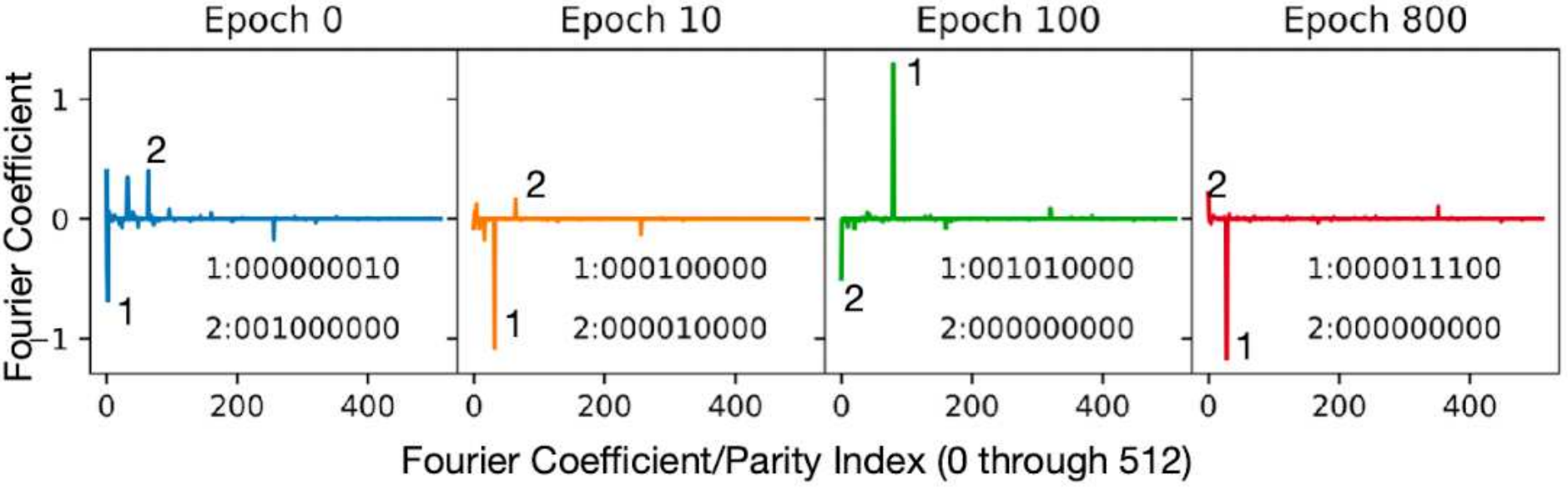}}\\
\subfloat[{\small FCs of block 2 after 4 training sessions with same initialization. The most dominant FC is stable across all runs.}]{\includegraphics[width=0.97 \columnwidth]{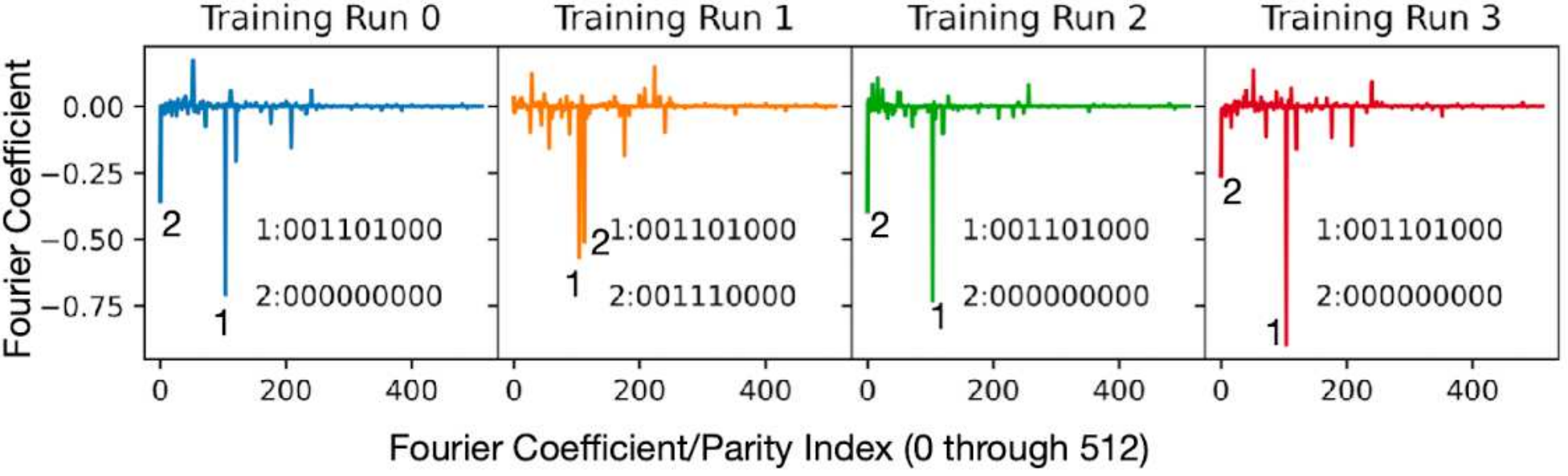}}
 \caption{\small Progression of TurboAE's Fourier Coefficients with Training. Similar behaviors seen for all the blocks. }
\label{fig:fcDynamics}
\vspace{-5mm}
\end{figure}

The trained TurboAE was found to have a few dominant Fourier coefficients~\cite{US-ISIT-2022, US-Allerton-2022}. One can hypothesize that this might be a general phenomenon when training this network. To investigate this question,
we trained TurboAE-binary  several times from scratch as described in \cite{turboAE}.
At convergence, the Fourier space appears to almost always be dominated by a few large FCs.
In the randomly selected examples in Fig. \ref{fig:fcDynamics}(a) 95\% of the total energy (sum of $\hat{f}^2(S)$)  is for at most 5 (out of 32) FCs. 

Furthermore we observed that at initialization, the dominant FCs almost always correspond to one bit parities. However, with training, higher degree parities emerge as dominant, see Fig. \ref{fig:fcDynamics}(b). Although the setups differ, these observations may be related to recently observed staircase properties \cite{abbe}.


We also trained TurboAE multiple times starting with the same initialization of the neural net weights to evaluate how stable the training process is at convergence. We found that it \emph{is} somewhat stable w.r.t. the dominant Fourier coefficients, as shown for some runs on Fig. \ref{fig:fcDynamics}(c). 
This begs further questions about the loss landscape of Turbo-like codes, which we propose to study also using a Fourier lens next.

\subsection{Local Optimality of parities for Turbo Codes}
The loss landscape of the TurboAE network is a function over 150,000
parameters, depending on the network. The network parameters of the encoder determine the FCs of the encoding function, and so local minima of the latter can be helpful for understanding the former.
Thus we study the loss landscape in terms of the Fourier parameterization with 512 parameters, fixing the decoder to BCJR. 

To study the loss landscape of generic \emph{non-recursive} Turbo codes in Fourier space, we constructed a parametric
Turbo code of block length $L$(=10), memory 4,  
parameterized not by CNNs, but by the FC of its constituent codes. 
It is not clear whether the encoder part of the TurboAE network can implement any triple of 5-variable Boolean functions, therefore the observations might not transfer directly to TurboAE.

For each block $b \in \left\{ 1, 2, 3\right\}$, use a pseudo-Boolean function  $f_{b,\Theta_b} :\{-1, 1\}^5 \rightarrow \mathbb{R}$
, as the constituent code $f_{b,\theta}(\cdot)$ of Fig. \ref{fig:encdec}, which is completely determined, hence parameterized by its FC  $\Theta_b = \hat{f}_b$. These form the 
 Turbo encoder $\mathcal{E}_\Theta $. We use a standard six iteration BCJR decoder for $\mathcal{E}_\Theta $ as the Decoder $\mathcal{D}_{\Theta }$. We use expected binary cross entropy ($BCE$) between the input and the decoded output as \emph{loss} $\mathcal{L}$, i.e.
 
$\mathcal{L}(\Theta) = \underset{\substack{ x \sim\mathcal{U}^L\\ {z {\sim \mathcal{N}^L}}}}{\mathbb{E}} \left[\frac{1}{L}{\underset{i \in [L]}{\sum}} BCE\left(x_i, \mathcal{D}_{\Theta }\left(\mathcal{E}_{\Theta}(x) + z\right)_i\right)\right]\quad.$

Here $z$ is the \emph{i.i.d.} noise sampled from AWGN channel i.e. $\mathcal{N}_{0,\sigma }$, where $\sigma $ corresponds to SNR = 1dB. We control the power by keeping the squared sum of the FCs to be 1, which constrains the average power of each bit to also be 1 due to Parseval's Theorem: 
$\underbrace{\underset{\mathbf x \sim \mathcal{U}^5}{\mathbb E}\left[f_b\left(\mathbf x\right)^2\right]}_{{\text{avg power}}} = \underset{S \subseteq [5]}{\sum_{}}{\hat{f}_b\left(S\right)^2}  = \left\|\Theta_b \right\|_2^2$.






Our hypothesis is that triples of different parity functions are all local minima, but there are other triples that are not local minima. We ran the following experiment, with results consistent with the hypothesis.

\begin{figure}
\centering
\subfloat[Line Joining two parities. Local minima at 0 and 1 show that both parities are locally optimal on this line.]{\includegraphics[width= 0.235\textwidth]{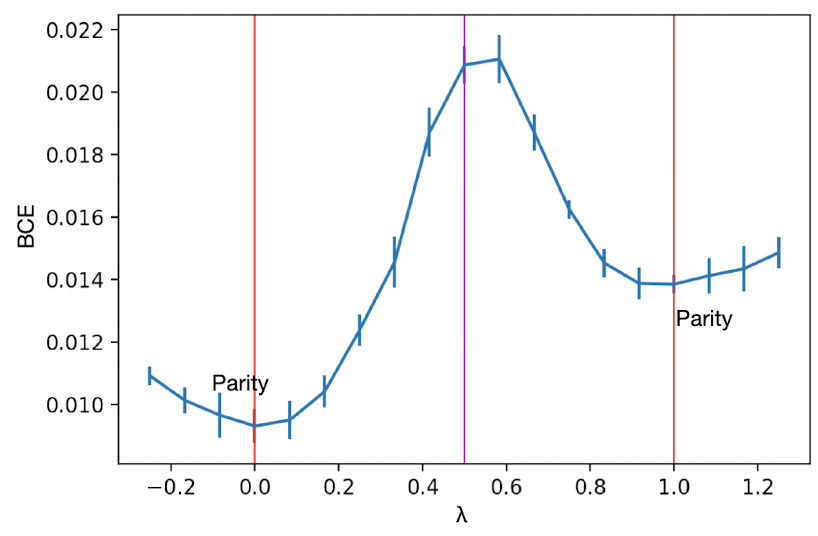}}
\hfill
\subfloat[Line Joining two non-parity Bent functions. Neither of the two are locally optimal.]{\includegraphics[width= 0.235\textwidth]{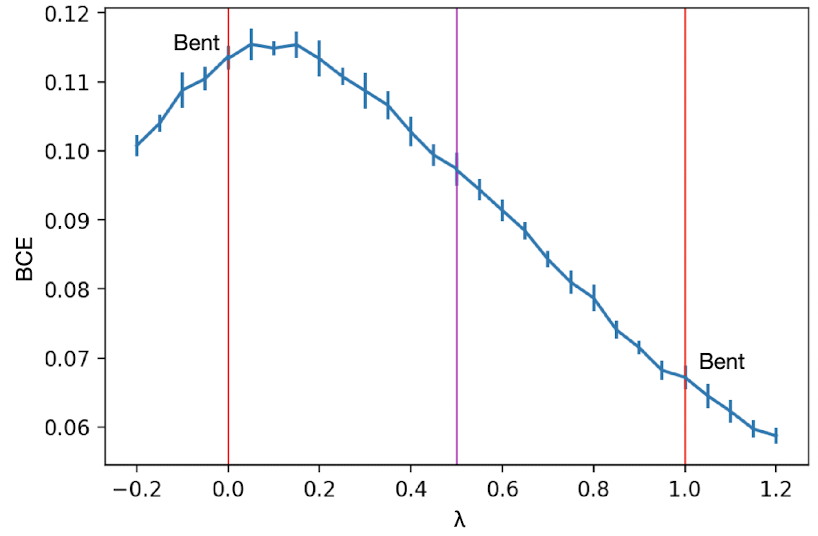}}
 \caption{\small BCE landscape on parametric line $\lambda \Theta'' + (1-\lambda )\Theta'$ joining combinations of parity/non-parity functions. $\lambda = 0,1$ correspond to $\Theta', \Theta''$ respectively.}
\label{fig:lossLandscape}
\vspace{-5mm}
\end{figure}

Pick $\Theta', \Theta''$ corresponding to triples of different parities $\chi_{\Theta'} = (\chi'_1, \chi'_2, \chi'_3)$ and $\chi_{\Theta''} = (\chi''_1, \chi''_2, \chi''_3)$ respectively. Evaluate $\mathcal{L}$ over several points on the line joining $\Theta'$ and $\Theta''$ with power re-normalization. 
Evaluating a point representing a pseudo-Boolean function involves running BCJR for that function and computing the BCE.
We found that $\Theta'$ and $\Theta''$ were always local optima on this line. On the other hand, 
the triple formed by three copies of the bent function $x_1 x_2 \oplus x_3 x_4$
on different subsets of 5 variables is not a local minimum.
The results are illustrated in
Fig.  \ref{fig:lossLandscape}, and Fig \ref{fig:lossLandscape2} in the Appendix.

\section{Training loss functions: BCE and BER}


We now investigate the implications of optimizing  BCE from theoretical and empirical perspectives. We consider optimizing an encoder function $f: \F_{2}^{k} \to S$ where $S \subset \R^n$ is bounded and $R = k/n$ is our code rate. We take $U \sim \text{Unif}[\F_{2}^{k}]$ and $Y \in \R^n$ to be random variables representing the input and received sequence, respectively. Note $Y$ depends on $f$. Our optimization problem is then:
\begin{problem}
Find encoder $f : {\F_{2}^{k}} \to S$ and soft decoder $g \in \R^n \to [0,1]^{k}$ 
that minimizes the expected BCE, $\C(f,g)$, where
\[\C(f,g) = \E\left[ \frac{1}{k} \sum\limits_{i=1}^{k}-U_{i} \lg g_{i}(Y) - (1-U_{i}) \lg (1-g_{i}(Y))]  \right].\]
\end{problem}

\subsection{Theoretical Analysis of BCE Minimization}
We re-write the BCE as 
\begin{equation}
\C(f,g) =  \frac{1}{k}\sum\limits_{i=1}^{k}\E[D_{KL}(\Prob[U_{i}=1|Y] || g_{i}(Y))] + \Ent(U_{i}|Y) 
\label{bce:decomp}
\end{equation}
where $\Ent(U_{i}|Y)$ is the conditional entropy
and $\E[D_{KL}(\Prob[U_{i}=1|Y]|| g_{i}(Y))$ is the expected Kullback-Leibler divergence
with the expectation taken over the distribution of received sequences.  
Note that for a fixed channel,  \(\Ent(U_i | Y)\) only depends on the encoder \(f\), and for a fixed encoder the KL-Divergence term $\E[D_{KL}(\Prob[U_{i}=1|Y]|| g_{i}(Y))$ only depends on the decoder, \(g\). We can easily establish the following proposition (proofs in Appendix):

\begin{proposition} \label{thm:decomp_minimizer}
Consider a fixed encoder $f: \F_{2}^k \to S$. 
The decoder $g: S \to [0,1]^k$ defined elementwise as $g_i(y) := \Prob(U_{i}=1|Y=y)$
is the unique a.s. minimizer of $\C(f, g)$. 
\end{proposition}

Note that this is exactly the soft-output MAP decoder, a minimizer of the BER $\B(f,g)$, for a fixed encoder $f$. This means that, given \(f\), minimizing BCE and BER both reduce to finding a soft-MAP decoder for \(f\). Thus we denote \(\C(f), \B(f)\) to be \(\C(f, g_{MAP(f)}), \B(f, g_{MAP(f)})\), where \(g_{MAP(f)}\) is the soft-MAP decoder for \(f\). Hence, finding a decoder that minimizes BCE finds a decoder that minimizes BER. The same does \textit{not} hold for the general problem of finding an encoder-decoder pair; a memoryless counter-example (asymmetric in the input, unlike the BSC or AWGN) is shown in the Appendix. The best bounds possible relating BER and BCE are the following.

\begin{proposition} \label{2_sided_bound}
Let \(\B_i, \C_i\) denote the BER and BCE respectively on the \(i^{\text{th}}\) input bit. Then for all choices of $f$ and for all $i \in [k]$,
$2\mathbb{B}_{i}(f) \leq \mathbb{C}_{i}(f) \leq \mathbb{H}_{2}(\mathbb{B}_{i}(f))$, and 
in particular
$$2\mathbb{B}(f) \leq \mathbb{C}(f) \leq \frac{1}{k} \sum\limits_{i=1}^{k} \mathbb{H}_{2}(\mathbb{B}_{i}(f))$$
where $\mathbb{H}_{2}(p)$ denotes the binary entropy function with parameter $p\in [0,1]$.
Furthermore, these bounds are tight in the sense that for any  BER \(t \in [0, \frac{1}{2}]\) and side of the bound, there exists a channel and an encoder which makes that side an equality.

\end{proposition}

\subsection{Empirical Application of BCE Decomposition} \label{section:training_encoders}

In light of equation~\eqref{bce:decomp}, it seems reasonable to decouple the optimization process: optimize the encoder conditional entropy first, then optimize a soft decoder to minimize its KL-divergence with respect to the good encoder. The major obstacle is estimating the conditional entropy of the encoder (or equivalently, estimating the KL-divergence of the decoder). 
We opt for a naive approach to get around this issue, but more sophisticated approaches can be found in \cite{Poczos_Schneider, Paninski_2003}.
If we were optimizing convolutional codes, we could take advantage of the fact that the BCJR \cite{bcjr1974} is a MAP decoder. However, in the case of general Turbo codes, we have no such efficient MAP decoder. To get around this we propose training a Turbo-like encoder (top of Fig. \ref{fig:encdec}, but allowing for real-valued outputs) at short block lengths, \(k=16\), using a brute-force marginalization MAP decoder. The encoder is trained with many choices of interleaver so it may generalize well at larger block lengths. Then, we train a neural decoder on the same encoder at our block length \(k = 100\). This is in direct contrast to the approach in \cite{turboAE}, where the authors alternated between training the encoder and decoder until they jointly converged. See Algorithm \ref{alg:alt_train} in the Appendix for a more precise formulation of the training procedure.

\subsubsection{Methods}

Like \cite{turboAE}, we train a non-systematic, non-recursive turbo code of rate $R = \frac{1}{3}$ for use at block length \(100\). 
Rather than training a neural encoder, we directly train the input-output table of a window \(w=5\) (memory \(4\)), possibly nonlinear, automata encoder, borrowing terminology from \cite{Bazzi_Mahdian_Spielman_2009}. That is, we parameterize our encoder by the outputs of a function \(h: \F_2^w \to \R^3\). This function is slid over the input bits in \(\F_2^k\) as in a non-recursive convolutional encoder. The third stream is convolved over an interleaved copy of the input instead. \(h\) is initialized with a normal distribution of mean 0 and variance 1. We also tried initializing \(h\) with a parity function, but found it to be at best as good as the normal initialization. Details can be found in the Appendix.



To enforce the power constraint 
we use a different method than in \cite{turboAE}. We instead analytically compute power using \(h\). We then center and rescale \(h\) after each gradient update so that its power is \(1\). Precise derivation of this power normalization can be found in the Appendix. This helped reduce much of the noise in the training process introduced by the power-normalization in \cite{turboAE}. This method could also be applied when training a neural encoder as in \cite{turboAE}. Once we had a trained encoder, we then trained a neural decoder at larger block length with the same architecture as in \cite{turboAE}.


\subsubsection{Results}

The encoder converges fairly quickly (200 steps). The training curve is included in the Appendix (Fig.~\ref{fig:enc_dec_training_curves}).
Over several training runs, we were unable to find an encoder with as low a conditional entropy as TurboAE-cont. The discrepancy may come: (1) our encoder was trained at a block length of 16, while TurboAE-cont at block length 100, and (2) TurboAE-cont was trained by alternating back-and-forth between optimizing the encoder and the decoder, which may have avoided local optima our training scheme runs into.


We show the evolution of the FC for our final encoder in Fig.~\ref{fig:fc_evolution}, and the coefficient evolution of another trained encoder in the Appendix. By the end, only a few dominant FC remain, echoing what we saw in Section~\ref{section:fc_dominance}. The FC change significantly from the initialization. 

\begin{figure}
    \centering
    \includegraphics[width=\columnwidth]{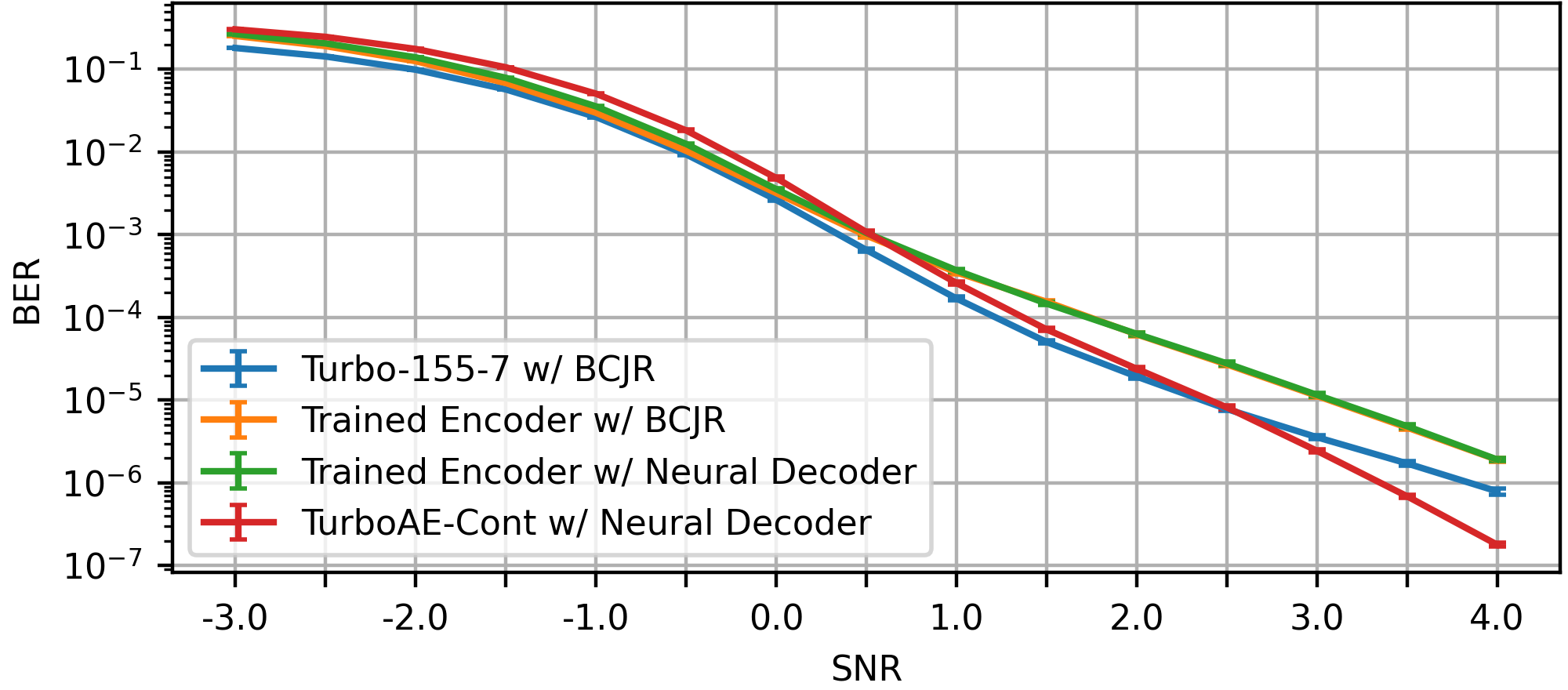}
    \caption{\small BER performance of (1) a benchmark RSC turbo code, (2)
    our trained encoder paired with BCJR, (3) our trained encoder paired with our neural decoder, (4) TurboAE-cont. TurboAE-cont tends to perform worse at SNRs below 0.5, while showing significant outperformance at SNRs above 3.0. 
    }
    \label{fig:snr_results}
    \vspace{-5mm}
\end{figure}

When training the decoder, our optimization proceeds relatively quickly compared to TurboAE-cont. See Fig.~\ref{fig:enc_dec_training_curves} in the Appendix for the training curve. The total number of steps required for our procedure is only 300,500 whereas TurboAE-cont requires 480,000. In Fig.~\ref{fig:snr_results} we see that the performance of our encoder-decoder is slightly worse than TurboAE-cont above SNR 1.0. 
Observe, if we replace our decoder with BCJR, the performance is almost exactly the same. The decoder may have learned a BCJR-like decoding algorithm which proved to be a local minimum. 
Nonetheless, our encoder-decoder pair suggests that our training scheme is a viable optimization strategy.



\begin{figure}
    \centering
    \includegraphics[width=\columnwidth]{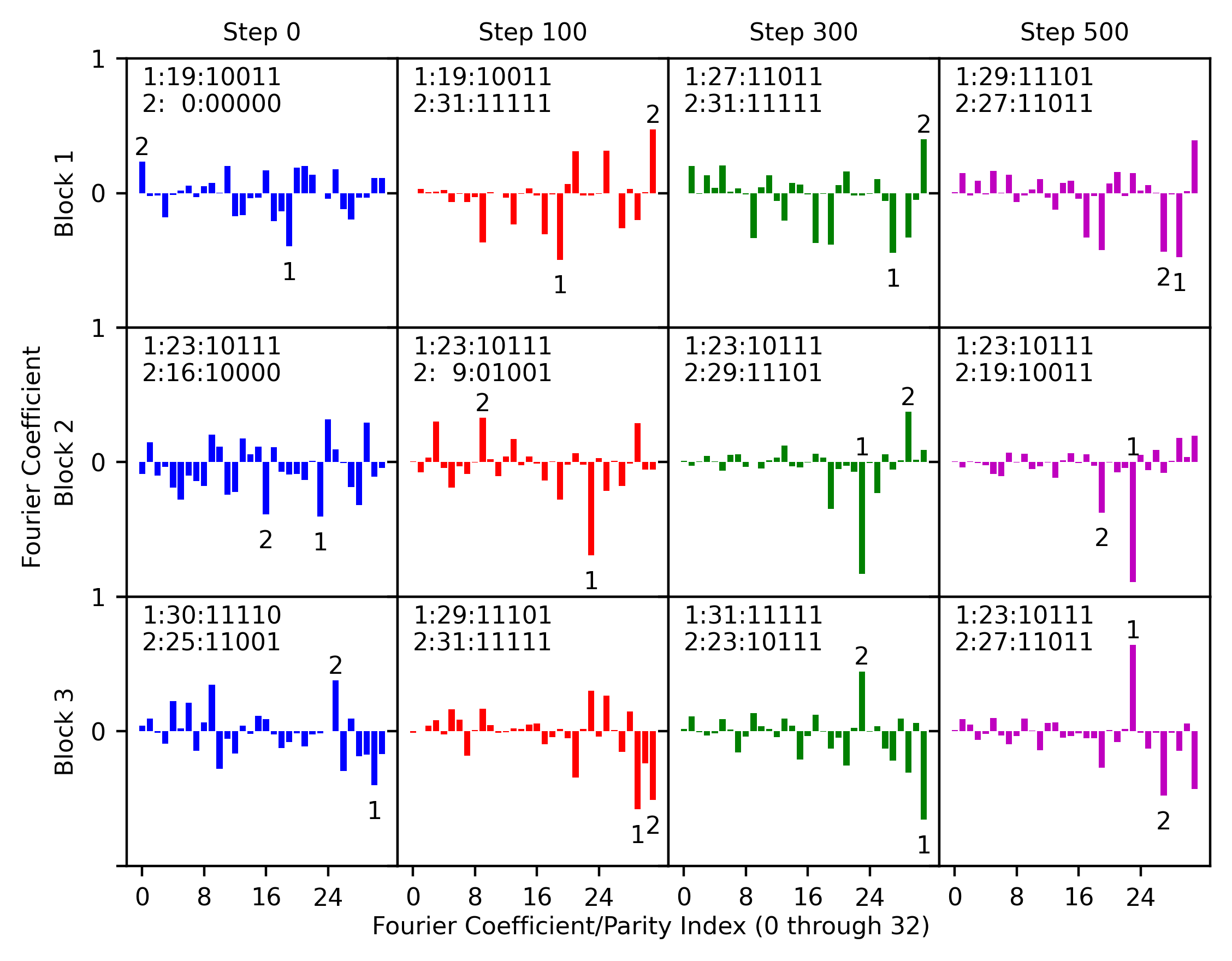}
    \caption{\small Evolution of FC during training of the encoder from Figure~\ref{fig:enc_dec_training_curves}. The largest coefficients are marked and their corresponding parity is annotated in the respective subplot. Note how a few dominant coefficients emerge and persist during training. 
    }
    \label{fig:fc_evolution}
    \vspace{-3mm}
\end{figure}


\section{Conclusions}

We presented several new tools which may help in the interpretation of training aspects of DL-ECCs, including 1) the application of the Goldreich-Levin algorithm to finding the best parity / linear  approximation to a black-box encoding function, where its efficiency is useful when there are a large number of input variables but perhaps a more local and/or sparse structure (outputs depend only on a few of those inputs each) of the final learned code; 2) the use of FC to understand the loss landscape when training DL-ECCs and also as a possible parameterization for learning codes; 3) observations relating the BCE and BER and a principled alternative approach for training /  optimizing DL-ECCs. 
While our experiments showed the viability of our alternate training scheme, there are still many more aspects to explore. In particular, with good estimation of the conditional entropy of an encoder at larger block lengths, we expect performance will improve.
In addition, the neural decoder of \cite{turboAE} was designed to mimic an iterative BCJR decoder. However, designing the neural network to mimic exact inference algorithms (e.g. a junction tree \cite{junction-tree-ref}) could lead to a better approximation of the MAP decoder.
From a bigger-picture perspective, we hope that this decomposition and the usage of Fourier coefficients both as an alternative representation and as a tool for understanding training, will lead to a more principled approach toward the training of deep-learned error-correcting codes. This could open doors to a more systematic way of finding such codes for different channels.

\bibliographystyle{IEEEtran}
\bibliography{aaai22, reference}

\appendices

\section{Interpretations of TurboAE-binary from \cite{US-ISIT-2022}}

\begin{table}[h]
\centering
\begin{tabular}{|p{1.1cm} | p{6cm} |} 
 \hline
 Block \# &  Approximate  expressions for output \#$i$ \\ [0.5ex] 
 \hline
 1 & $1 \oplus u_1 \oplus {u}_2 \oplus u_3 \oplus {u}_4 \oplus u_5$ \\
\hline
 2 & $u_1 \oplus u_3 \oplus u_4 \oplus u_5 $  \\
 \hline
 3 & \text{Solution 1: }$u_1 \oplus u_2 \oplus u_4$\par 
  \text{Solution 2: } $1\oplus u_1 \oplus u_2 \oplus u_3 \oplus u_4$\par
  \text{Solution 3: } $1 \oplus u_1 \oplus u_2 \oplus u_4 \oplus u_5$\par
  \text{Solution 4: } $1\oplus u_1 \oplus u_2 \oplus u_3 \oplus u_4 \oplus u_5$\\
  \hline
  where & $u_1 = x_{i+2}$, $u_2 = x_{i+1}$, $u_3 = x_j$, \par $u_4 = x_{i-1}$, $u_5 = x_{i-2}$, $x = \text{inputs}$ \\
  \hline
\end{tabular}
\caption{\small Best affine approximations for TurboAE-binary's encoder. Block 3 has four equally good approximations, from \cite{US-ISIT-2022}.}
\label{table:bestLinearApprox}
\end{table}

\begin{table}[h]
\centering
\begin{tabular}{|p{1.1cm} | p{6cm} |} 
 \hline
 Block \# & Expression for output \#$i \in \{3, 4, \cdots, 98\}$ \\ [0.5ex] 
 \hline
 1 & $1 \oplus u_1 \oplus \bar{u}_2 \oplus u_3 \oplus \bar{u}_4 \oplus u_5$ \par
 $\oplus \ \bar{u}_2 u_3 \bar{u}_4 \oplus u_1 \bar{u}_2 u_3 \bar{u}_4 u_5$ \\
 2 & $u_1 \oplus u_3 \oplus u_4 \oplus u_5 $  \\
 3 & $u_1 \oplus u_2 \oplus u_4 \oplus \bar{u}_3 \bar{u}_5$ \\
  \hline
  where & $u_1 = x_{i+2}$, $u_2 = x_{i+1}$, $u_3 = x_i$, \par $u_4 = x_{i-1}$, $u_5 = x_{i-2}$, $x = \text{inputs}$ \\
  \hline
\end{tabular}
\caption{\small Exact expressions for TurboAE-binary encoder's non-boundary bits, taken from \cite{US-ISIT-2022}.}
\label{table:exact}
\end{table}

\section{Implementation of the Goldreich-Levin algorithm}

The Goldreich-Levin algorithm aims to find sets $S$ with Fourier coefficient magnitude $|\hat{f}(S)|>\gamma$.
The details of the Goldreich-Levin algorithm are shown in Algorithm \ref{gl} \cite{o2014analysis}.




\begin{algorithm} 
	\caption{Goldreich-Levin Algorithm (length of input sequence $n$,  query function $f$, threshold $\gamma$, confidence $\delta$ )} 
	\label{gl} 
	\begin{algorithmic}[1]
		\STATE Initialization: $k = 0$, $U = \emptyset$
		\STATE Randomly generate a list $L\gets (*,*,\cdots,*)$, which is a collection of sets such that $L = \{U\cup T: T\subseteq[n]\setminus[k]\}$
            \FOR {each $k \in [1,n]$}
                \FOR {each $B\in L$, $B = (a_1, \cdots,a_{k-1},*,\cdots,*)$}
                \STATE Let $B_{a_k} = (a_1, \cdots,a_{k},*,\cdots,*)$ for $a_k = 0,1$
                \STATE Estimate the Fourier weight $W(B_{a_k})$ within $\pm \frac{\gamma^2}{4}$ with probability at least $1-\delta$
                \STATE Remove $B$ from $L$
                \STATE Add $B_{a_k}$ to L if the estimated weight $W(B_{a_k})\geq \frac{\gamma^2}{2}$
                \ENDFOR
            \ENDFOR
            
        \STATE Output: L
	\end{algorithmic} 
\end{algorithm}

To run this algorithm in practice, one needs to know $\gamma$ and ensure that enough queries are made. In our application, we simply want to find the set with largest Fourier coefficient magnitude (or multiple sets if they all have roughly equal Fourier coefficient magnitudes). Thus, there is no prior-fixed $\gamma$ in existing applications. We detail how we  experimentally selected $\gamma$ and the number of queries here without prior knowledge.



The problem is one needs to jointly select $\gamma$ and the number of queries. We could envision using a binary search for $\gamma$, where for each $\gamma$ we start with a small number of queries and double it until we have a low variance in the output sets (which are random given the randomzied nature of the algorithm). 

\begin{figure}[h]
     \includegraphics[width=\columnwidth]{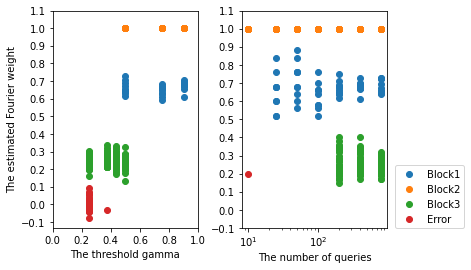}
     \caption{\small Estimated Fourier weights of Goldreich-Levin of three TurboAE blocks as a function of $\gamma$ (Left) and the convergence of Fourier weight over the number of queries (Right).}
     \label{fig:gltrain}
\end{figure}

What we did in practice here was the following heuristic approach:
we first fix the number of queries to a reasonably large number; in our experiments we took 800 queries (still small compared to the $2^{100}$ input space).  \footnote{By \cite[Problem 1.5]{o2014analysis}, there can only be at most 1 Fourier coefficient with magnitude above 0.5.} We first set $\gamma=0.5$ and ran GL to see whether a single set has a large Fourier coefficient. 
In our example, for $\gamma > 0.5$, here we test $\gamma \in \{0.75, 0.9\}$, only Block 1 and Block 2 have stable outputs (stable means the same output set is consistently produced when the randomized GL runs several times with different initializations). This indicates there is a single dominant parity. If the algorithm does not return a result for $\gamma > 0.5$, the $\gamma$ is lowered and a binary search on $\gamma\in (0,0.5)$ is used to find the largest $\gamma$ that produces a stable output set. In our example, Block 3 does not have one dominant parity, but rather has 4.  In running the binary search, we regard  any re-estimated Fourier weight less than $\frac{\gamma^2}{2}$ as an error and an indication that the $\gamma$ is too small (indicated by red dots).  Taking threshold $\gamma \in \{0.25, 0.375, 0.4375\}$, from the Fig. \ref{fig:gltrain} (Left), it can be seen that the output is not reliable until $0.4375$. Therefore, in this work, we choose $\gamma_1 = 0.8$ for Block 1, $\gamma_2 = 0.9$ for Block 2, and $\gamma_3 = 0.45$ for Block 3. 


To investigate the effect of the number of queries on convergence of the Fourier weight, we test the Goldreich-Levin algorithm on different numbers of queries (10, 25, 50, 100, 200, 400, 800), and show 10 runs for each. In Fig. \ref{fig:gltrain} (Right) we plot the weights of the found sets and see that Block 3 do not have output lists when the number of queries is small, Block 2 converges quickly, and Block 1 roughly converges after 200 queries. When the number of queries is too small, there is an error output list for Block 2 (indicated by red dots).
Block 2's rapid convergence is likely due to the fact that the true function is a parity and hence has one large Fourier coefficient. That the others converge more slowly is likely due to the fact that their exact representations seen in Table \ref{table:exact} are more non-linear. Block 3 has several equally large coefficients.



\section{Loss landscape plots}

See Fig. \ref{fig:lossLandscape}, referred to in Section \ref{sec:losslandscape}.    


\begin{figure*}
\centering
\subfloat[$P,P' = (1, 10, 21),(1, 10, 23)$]{\includegraphics[width= 0.3\textwidth]{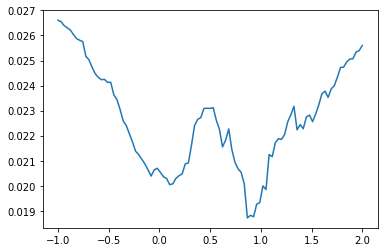}}
\subfloat[$P,P' = (1, 10, 23),(1, 10, 22)$]{\includegraphics[width= 0.3\textwidth]{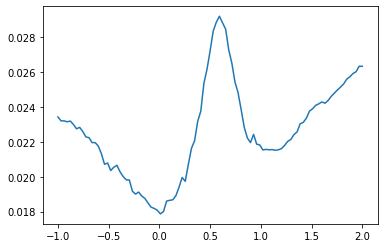}}
 \caption{\small Binary Cross Entropy landscape on parametric line $\lambda \Theta'' + (1-\lambda )\Theta'$ joining two parity combinations $\Theta'$ and $\Theta''$. $\lambda = 0,1$ correspond to $\Theta', \Theta''$ respectively. Local minimums around $\lambda = 0,1$ suggest that $\Theta', \Theta''$ are locally optimal codes on the line joining $\Theta',\Theta''$.}

\label{fig:lossLandscape2}
\end{figure*}

\section{Proof of Proposition \ref{thm:decomp_minimizer}}

\begin{proof}
Recall equation~\eqref{bce:decomp}:
\begin{equation}
\C(f,g) =  \frac{1}{k}\sum\limits_{i=1}^{k}\E\left[D_{KL}(\Prob[U_{i}=1|Y] || g_{i}(Y))\right] + \Ent(U_{i}|Y).
\end{equation}
For a fixed \(f\), our choice of \(g\) does not affect \(\Ent(U_i|Y)\), so it only matters how it affects \(\E\left[D_{KL}(\Prob[U_{i}=1|Y] || g_{i}(Y))\right]\). Recall by Gibb's Inequality that KL-Divergence is always nonnegative, and is 0 if and only if \(\Prob[U_{i}=1|Y=\mathbf{y}] = g_{i}(\mathbf{y})\). Thus, by defining \(\Prob[U_{i}=1|Y=\mathbf{y}] =: g_{i}(\mathbf{y})\) for each \(i \in [k]\), we ensure \(\frac{1}{k}\sum\limits_{i=1}^{k}\E\left[D_{KL}(\Prob[U_{i}=1|Y] || g_{i}(Y))\right] = 0\), its minimum possible value, and Gibb's Inequality ensures that this is the almost surely unique choice of minimizing \(g\).
\end{proof}

\section{Proof of 2-Sided Bound on BCE by BER \ref{2_sided_bound}}

For the proofs below, we take as definition for BER of an encoder-decoder pair $f,g$:

\[\B(f, g) = \E\left[\frac{1}{k} \sum_{i=1}^{k} U_i \1[g(Y)_i \leq \frac{1}{2}] + (1-U_i) \1[g(Y)_i 
> \frac{1}{2}] \right]. \]
When $g$ is the soft-MAP decoder for $f$, this reduces to
\[\B(f) =  \frac{1}{k}\sum\limits_{i=1}^{k}\E\Big[\min(\Prob(U_{i}=1|Y), \Prob(U_{i}=0|Y))\Big].\]

The best bounds relating BER and BCE we can hope for are in Prop. \ref{2_sided_bound}, whose proof is the following:

\begin{proof}

Let $i \in [k]$ and fix $f$. Let $T = \min(\mathbb{P}(U_{i}=1|Y), \mathbb{P}(U_i=0|Y))$. Then 
\begin{align*}
&\mathbb{B}_{i}(f) = \mathbb{E}[T], \;\; \mathbb{C}_{i}(f) = \mathbb{H}(U_{i}|Y) = \mathbb{E}[\mathbb{H}_2(T)]
\end{align*}
Recall that entropy is a concave function, so all line segments connecting two points of $\mathbb{H}_2$ in $\left[0,1\right]$ lie in the undergraph of $\mathbb{H}_2$. Since $\mathbb{H}_{2}(0)= 0$ and $\mathbb{H}_{2}\left(\frac{1}{2}\right) = 1$, we know that $\mathbb{H}_{2}(s) \geq 2s$ $\forall s \in \left[0, \frac{1}{2}\right]$. Thus, by monotonicity of expecation, we have that
$\mathbb{C}_{i}(f) = \mathbb{E}[\mathbb{H}_{2}(T)]\geq 2\mathbb{E}[T] = 2\mathbb{B}_{i}(f)$
proving the LHS inequality.
For the second, note again that entropy is a concave function, so we can apply Jensen's Inequality. Then
$\mathbb{C}_{i}(f) = \mathbb{E}[\mathbb{H}_{2}(T)]\leq \mathbb{H}_{2}(\mathbb{E}[T]) = \mathbb{H}_{2}(\mathbb{B}_{i}(f))$
giving us the RHS inequality.
The second inequality follows by averaging together the inequalities for each component.
\end{proof}

\subsection{Tightness of 2-sided bound}

We will show that $\forall t \in  [0, 1/2]$, for each of the statements below, there exist encoders $f: \F_2^n \to S$ and a noise models on $S$ so that
\begin{enumerate}
    \item $2 \B_i(f) = \C_i(f) = 2t$
    \item $\C_i(f) = \Ent_2(\B_i(f)) = \Ent_2(t)$
\end{enumerate}
This will establish that our bound from Prop.~\ref{2_sided_bound} is tight. Note that here $\B_i$ and $\C_i$ are functions of \emph{both} $f$ and our noise model.

\subsubsection{Upper bound is tight}

\begin{proof}
Take $f: \F_2 \to \F_2$ to be the identity function, and let
\[P := \begin{pmatrix} t & 1-t \\ 1-t & t\end{pmatrix},\]
be the transition matrix of our channel (a binary symmetric channel).
Then it may be shown that $\B(f) = t$ and \(\C(f) = \Ent(U|Y) = \Ent_2(t) = \Ent_2(\B(f))\), proving tightness of the upper bound.

\end{proof}

\subsubsection{Lower bound is tight}

\begin{proof}
Take \(f: \F_2 \to [3]\) to be any injective encoder. We  specify our noise model by specifying $\Prob(U=0|Y=i)$ and $\Prob(Y=i)$, then use these to compute the symbol transition probabilities. Denote $\Prob(U=0|Y=i) =: x_i$ and $\Prob(Y=i) =: y_i$. Then our constraints are 
\[
\begin{aligned}
\sum_{i=1}^3 y_i x_i &= \Prob(U=0) = \frac{1}{2}, \,\, \sum_{i=1}^3 y_i &= 1,
\end{aligned}
\]
and 
\[
\begin{aligned}
\B(f) &= \sum_{i=1}^3 y_i \min(x_i, 1 - x_i) = t \\
\C(f) &= \sum_{i=1}^3 y_i \Ent_2(x_i) = 2t,
\end{aligned}
\]
 where the last constraint enforces that our lower bound holds with equality. Consider $x_1 = 1$, $x_2 = 0$, $x_3 = 1/2$. Let $y_3 = 2t$ and $y_1 = y_2 = 1/2- t$. Since $t \in [0, 1/2]$ this makes sense.  One can verify that our constraints are met.

Using $y$ and $x$ we explicitly construct our noise model in terms of symbol transition probabilities. Denote $p_{i,j} := \Prob(Y=i | U=j)$. By Bayes' Theorem
\[
\begin{aligned}
    p_{i,0} &= \frac{\Prob(U=0|Y=i)\Prob(Y=i)}{\Prob(U=j)} = 2x_iy_i
\end{aligned}
\]
and
\[
\begin{aligned}
    p_{i,1} &= \frac{\Prob(U=1|Y=i)\Prob(Y=i)}{\Prob(U=j)}= 2(1-x_i)y_i.
\end{aligned}
\]
Substituting, our symbol transition probabilities can be expressed as
\begin{align}
    P& := \begin{pmatrix} p_{1,0} & p_{1,1} \\ p_{2,0} & p_{2,1} \\ p_{3,0} & p_{3,1} \end{pmatrix}
    = \begin{pmatrix} 1-2t & 0 \\ 0 & 1-2t \\ 2t & 2t \end{pmatrix}
\end{align}

Injectivity of $f$ was used to ensure we can assign different symbol transition probabilities for different values of $U\in\F_2$.

\end{proof}

\section{Counterexample that shows BER (and BLER) minimizing encoders are not BCE minimizers} \label{ce_min:counterexample}

\begin{proposition} \label{thm:ce_min_counterexample}
There exists a channel so that all encoder minimizers of cross-entropy are not encoder minimizers of the bit error rate. 
\end{proposition}

To show this, we will take the following parameters
\begin{enumerate}
    \item $k=1$, so $U \sim \text{Unif}[\F_2]$. In this case BER = BLER.
    \item Encoder $f : \F_2 \to [4]$. Denote $X := f(U)$. This is a random variable.
    \item Random variable $Y \in [4]$ represents the corrupted channel output.
\end{enumerate}
In our one-bit case, we have 
\begin{equation}  \label{eq:cx_ber}
    \B(f) = \E \Big[ \min(\Prob(U=1 | Y), \Prob(U=0 | Y)) \Big]
\end{equation}
\begin{equation}  \label{eq:cx_ce}
    \C(f) = \Ent(U|Y) = \E \Big[ \Ent_2(\Prob(U=1|Y)) \Big].
\end{equation}
Our channel is represented as
\[P = \begin{pmatrix}p_{1,1}&p_{1,2}&p_{1,3}&p_{1,4} \\ p_{2,1}&p_{2,2}&p_{2,3}&p_{2,4} \\ p_{3,1}&p_{3,2}&p_{3,3}&p_{3,4} \\ p_{4,1}&p_{4,2}&p_{4,3}&p_{4,4}\end{pmatrix}\]
where $p_{ij}$ for $i, j \in  [4]$ represents $\Prob(Y=i|X=j)$. 
Using $\Prob(U=1) = \Prob(U=0) = 1/2$ and Bayes' rule, 
\[
\begin{aligned}
    \B(f) &= \sum_{i=1}^4 \frac{p_{if(1)} + p_{if(0)}}{2} \frac{\min(p_{if(1)}, p_{if(0)})}{p_{if(1)} + p_{if(0)}}\\
    &= \frac{1}{2}\sum_{i=1}^4 \min(p_{if(1)}, p_{if(0)})    
\end{aligned}
\]
and $\C(f)$ in terms of $P$ as:
\[
\begin{aligned}
    \C(f) &=  \E \Big[ \Ent_2(\Prob(U=1|Y)) \Big]\\
    &= -\sum_{i=1}^4 \frac{1}{2} p_{if(1)} \lg\left(\frac{p_{if(1)}}{p_{if(1)} + p_{if(0)}}\right) \\ 
    &+ \frac{1}{2} p_{if(0)} \lg\left(\frac{p_{if(0)}}{p_{if(1)} + p_{if(0)}}\right) \\
    &=-\frac{1}{2} \sum_{i=1}^4 p_{if(1)} \lg p_{if(1)} +  p_{if(0)} \lg p_{if(0)}  \\
    &- (p_{if(1)} + p_{if(0)}) \lg(p_{if(1)} + p_{if(0)}). \\
\end{aligned}
\]


For explicit construction of the counterexample we take advantage of the fact that $\min(x, 1-x)$ and $\Ent_2(x)$ weight $[0, \frac{1}{2}]$ differently. For $\min(x, 1-x)$, an improvement of $t$ on one input and a worsening by $t$ on another cancel each other out no matter what the original $x$ was. On the other hand, because of the curvature $\Ent_2(x)$ the net result depends on the value of $x$. Let our transition matrix be
\[P = \begin{pmatrix}
    0.24   & 0.15   & 0.24   & 0.056 \\
    0.26   & 0.15   & 0.26   & 0.343 \\
    0.2605 & 0.35   & 0.2605 & 0.25  \\
    0.2395 & 0.35   & 0.2395 & 0.35  \\
\end{pmatrix}\]
Then we try each possible $f$, representing each $f$ as $[f(0), f(1)]$:
\begin{center}
\begin{tabular}{|c|c|c|} 
 \hline
    \(f\) & \(\B(f)\) & \(\C(f)\) \\ [0.5ex] \hline
    \([1, 2]\) & 0.4     & 0.969 \\ \hline
    \([1, 3]\) & 0.5     & 1.0 \\ \hline
    \([1, 4]\) & 0.40275 & 0.943 \\ \hline
    \([2, 3]\) & 0.4     & 0.969 \\ \hline
    \([2, 4]\) & 0.40300 & 0.949 \\ [1ex] \hline
    \([3, 4]\) & 0.40275 & 0.943 \\ \hline
\end{tabular}
\end{center}
It is clear that $[1,2], [2,3]$ are minima of the BER with value $0.4$, while $[1,4],[3,4]$ are minima of cross-entropy with value $0.943$. Thus, they are not minimized by the same encoder, decoder pair. $\qed$

\section{New proposed training algorithm for TurboAE-like codes}
See algorithm~\ref{alg:alt_train} as referred to in Section~\ref{section:training_encoders}.

\begin{algorithm} 
	\caption{Alternative Training Scheme($\Theta_{ENC}$, $\Theta_{DEC}$, $k_{ENC} = 16$, $k_{DEC} = 100$, $s_{ENC}=500$, $s_{DEC}^{(1)}=150,000$, $s_{DEC}^{(2)}=150,000$)} 
        \label{alg:alt_train}
	\begin{algorithmic}[1]
		\STATE Randomly initialize encoder parameters $\Theta_{ENC}$
		\FOR {steps $1$ to $s_{ENC}$}
                \STATE Sample a batch of inputs, noise, and sample an interleaver
                \STATE Empirically estimate the cost $\frac{1}{k_{ENC}} \sum_{i=1}^{k_{ENC}} \Ent(U_{i}|Y)$ for $\Theta_{ENC}$ using the batch
                \STATE Update $\Theta_{ENC}$ using the gradient of the cost
                \STATE Enforce power constraint on $\Theta_{ENC}$ using algorithm \ref{alg:power_constraint}
            \ENDFOR
            \FOR {steps $1$ to $s_{DEC}^{(1)}$}
                \STATE Enforce power constraint on $\Theta_{ENC}$ for new block length \(k_{DEC}\) using algorithm \ref{alg:power_constraint}
                \STATE Sample a batch of inputs, noise, and interleavers
                \STATE Empirically estimate the $\C(\Theta_{ENC}, \Theta_{DEC})$ using the batch at block length $k_{DEC}$
                \STATE Update $\Theta_{DEC}$ using the gradient of the cost
            \ENDFOR
            \STATE Fix a randomly chosen interleaver \(\pi\) for further training
            \FOR {steps $1$ to $s_{DEC}^{(2)}$}
                \STATE Sample a batch of inputs and noise
                \STATE Empirically estimate the $\C(\Theta_{ENC}, \Theta_{DEC})$ using the batch at block length $k_{DEC}$ and fixed interleaver \(\pi\)
                \STATE Update $\Theta_{DEC}$ using the gradient of the cost
            \ENDFOR
            
        \STATE return($\Theta_{ENC}$, $\Theta_{DEC}$) 
	\end{algorithmic} 
\end{algorithm}

\section{Analytic Derivation of Power Constraint}

We are given a nonrecursive Turbo code, \(f: \F_2^k \to \R^{1/R \times k}\), of window size \(w\) (memory \(w-1\)) and interleaver $\pi: [k] \to [k]$ for input size $k$ and rate $R = k/n$. This code is parameterized by $1/R$ generating functions \(h^{(s)}: \F_2^w \to \R\), with \(s \in [1/R]\) in the following sense for input \(\mathbf{u} \in \F_2^k\) and \(i \in [k]\):
\[
\begin{aligned}
    f^{(s)}_i(\mathbf{u}) &= h^{(s)}(\mathbf{u}_{i-w+1 : i}) \qquad s \in [1/R - 1] \\
    f^{(s)}_i(\mathbf{u}) &= h^{(s)}(\pi(\mathbf{u})_{i-w+1 : i}) \qquad s = 1/R \\
\end{aligned}
\]
where \(\mathbf{u}_j = 0\) for \(j \leq 0\). From here we can directly compute the power of a code:
\[
\begin{aligned}
    &\E \left[ \frac{R}{k}\sum_{s=1}^{1/R} \sum_{i=1}^k f^{(s)}_{i}(\mathbf{u})^2 \right] \\ 
    &=\frac{R}{k} \sum_{s=1}^{1/R} \sum_{i=1}^k \E \left[ f^{(s)}_{i}(\mathbf{u})^2 \right] \\
    &= \sum_{s=1}^{1/R-1} \sum_{i=1}^k \frac{R}{k} \E \left[ h^{(s)}(\mathbf{u}_{i-w+1 : i})^2 \right] \\
    &\;\;\;\; + \sum_{i=1}^k \frac{R}{k} \E \left[ h^{(1/R)}(\pi(\mathbf{u})_{i-w+1 : i})^2 \right]  \\
\end{aligned}
\]
Taking into account that the each \(\mathbf{u}_i \sim \text{Unif}[\F_2]\) iid, we can simplify the above expression and split into boundary terms + main sequence terms:
\begin{align} \label{eq:pow_from_table}
   & \frac{R}{k} \sum_{s=1}^{1/R} \sum_{i=1}^{w-1} \E \left[ h^{(s)}(\mathbf{u}_{i - w + 1:i})^2 \right] \\
    & \;\;\;\; + \frac{R}{k} (k -w + 1) \sum_{s=1}^{1/R} \E \left[ h^{(s)}(\mathbf{u}_{1:w})^2 \right],
\end{align}
which can be directly computed for small \(w\).

To meet the power constraint we can simply rescale $h$ so equation~\eqref{eq:pow_from_table} evaluates to \(1\). However, rescaling by a larger constant lowers the effective SNR of the code. Note that for the AWGN channel, performance is determined by the relative arrangement of the codewords in Euclidean space. Thus, we wish to find a translation constant \(C\) that minimizes the rescaling constant \(S\) needed so \((h - C) / S\) produces an encoder with power 1. Observe that
\[
\begin{aligned}
    &\frac{R}{k}\sum_{s=1}^{1/R} \sum_{i=1}^k \E \left[ (f^{(s)}_{i}(\mathbf{u})  - C)^2 \right] \\ 
    &= \frac{R}{k}\sum_{s=1}^{1/R} \sum_{i=1}^k \E \left[ (f^{(s)}_{i}(\mathbf{u})^2 \right] - 2C \E \left[ f^{(s)}_{i}(\mathbf{u}) \right] + C^2\\ 
    &= C^2 - 2 C \frac{R}{k}\sum_{s=1}^{1/R} \sum_{i=1}^k \E \left[ f^{(s)}_{i}(\mathbf{u}) \right] + \cdots
\end{aligned}
\]
where the remaining terms are not relevant to finding the optimum value of $C$. Note that $C$ is minimized when 
\begin{equation} \label{eq:opt_center}
    C = \frac{R}{k}\sum_{s=1}^{1/R} \sum_{i=1}^k \E \left[ f^{(s)}_{i}(\mathbf{u}) \right]    
\end{equation}
which we can directly compute from \(h\) in similar fashion to \eqref{eq:pow_from_table}. This gives us algorithm~\ref{alg:power_constraint} for optimally enforcing the power constraint during training.

\begin{algorithm} 
	\caption{Constrain Power(generator $h$, block length $k$)} 
        \label{alg:power_constraint} 
	\begin{algorithmic}[1] 
		\STATE Explicitly calculate equation \eqref{eq:opt_center} using $h$ and block length $k$, and assign to $C$
		\STATE Explicitly calculate equation \eqref{eq:pow_from_table} using $h - C$ and block length $k$, and assign to $S$
            \STATE \(h \leftarrow \frac{h - C}{S}\)
            
        \STATE return(\(h\)) 
        \end{algorithmic} 
\end{algorithm}

\section{Additional Initialization tried for Encoder}

\begin{figure}[h]
     \includegraphics[width=\columnwidth]{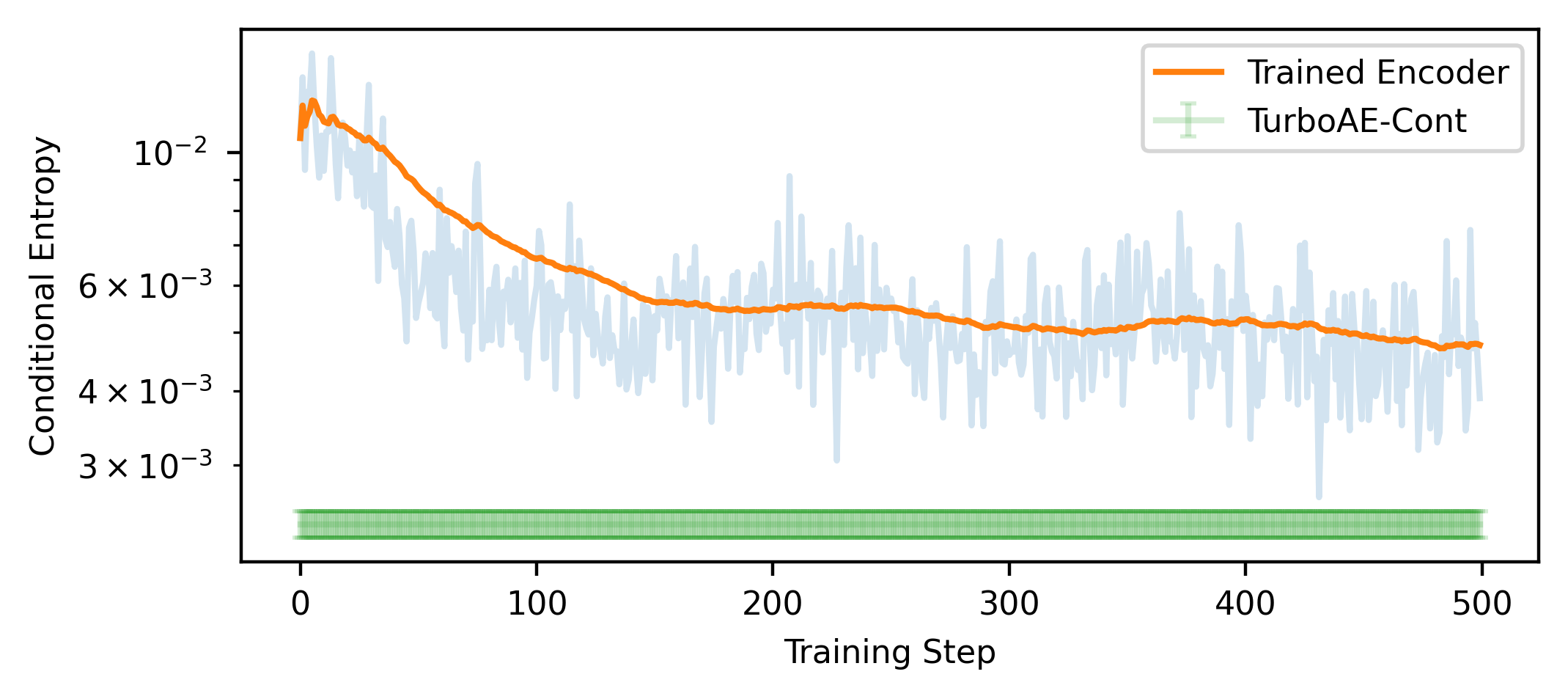}
     \caption{\small Training curve for a parity initialized encoder trained with conditional entropy at block length 16. The green bar shows the estimated conditional entropy of TurboAE-cont at the same block length. Its quick descent shows it is not a local optimum.}
     \label{fig:parity_training_curve}
\end{figure}

\begin{figure}[h]
     \includegraphics[width=\columnwidth]{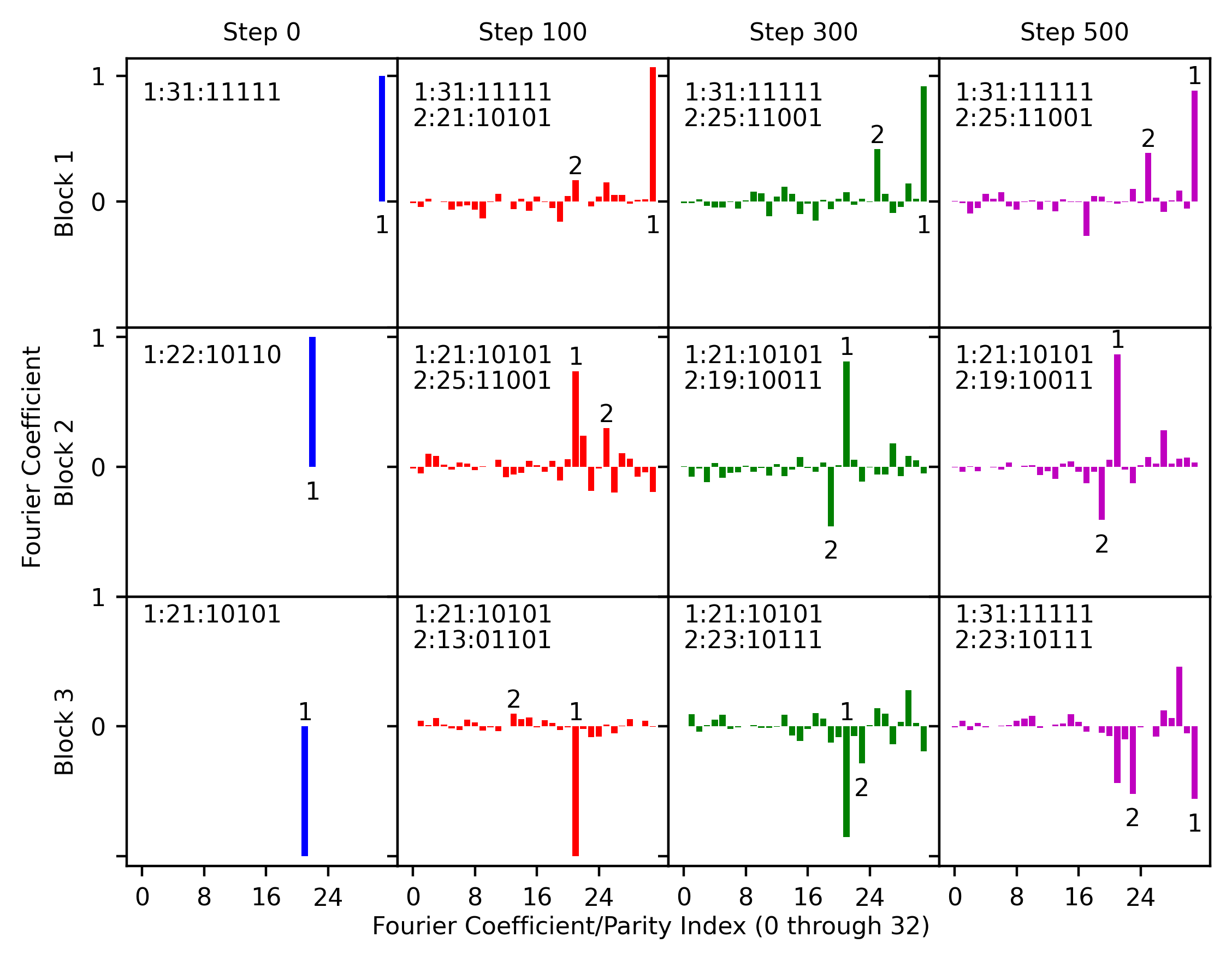}
     \caption{\small FC evolution during training of the parity initialized encoder from Fig.~\ref{fig:parity_training_curve}. As a parity, it starts with 1 coefficient, but during training it develops other dominant FC.}
     \label{fig:parity_fc_training}
\end{figure}

In addition to the initialization for \(h\) described in section \ref{section:training_encoders}, we also tried initializing \(h\) with a parity. That is, \(h\) was initialized with a uniformly random chosen parity function from \(\F_2^w \to \F_2^3\) in which the \(w^{\text{th}}\) input bit has nonzero influence on all 3 outputs. We show the training curve (Fig.~\ref{fig:parity_training_curve}) and FC evolution (Fig.~\ref{fig:parity_fc_training}) for one of our runs. The example shows that parities are not always local optima when training with conditional entropy.

\section{Training Curves for Encoder and Decoder}
\begin{figure}
    \centering
    \includegraphics[width=\columnwidth]{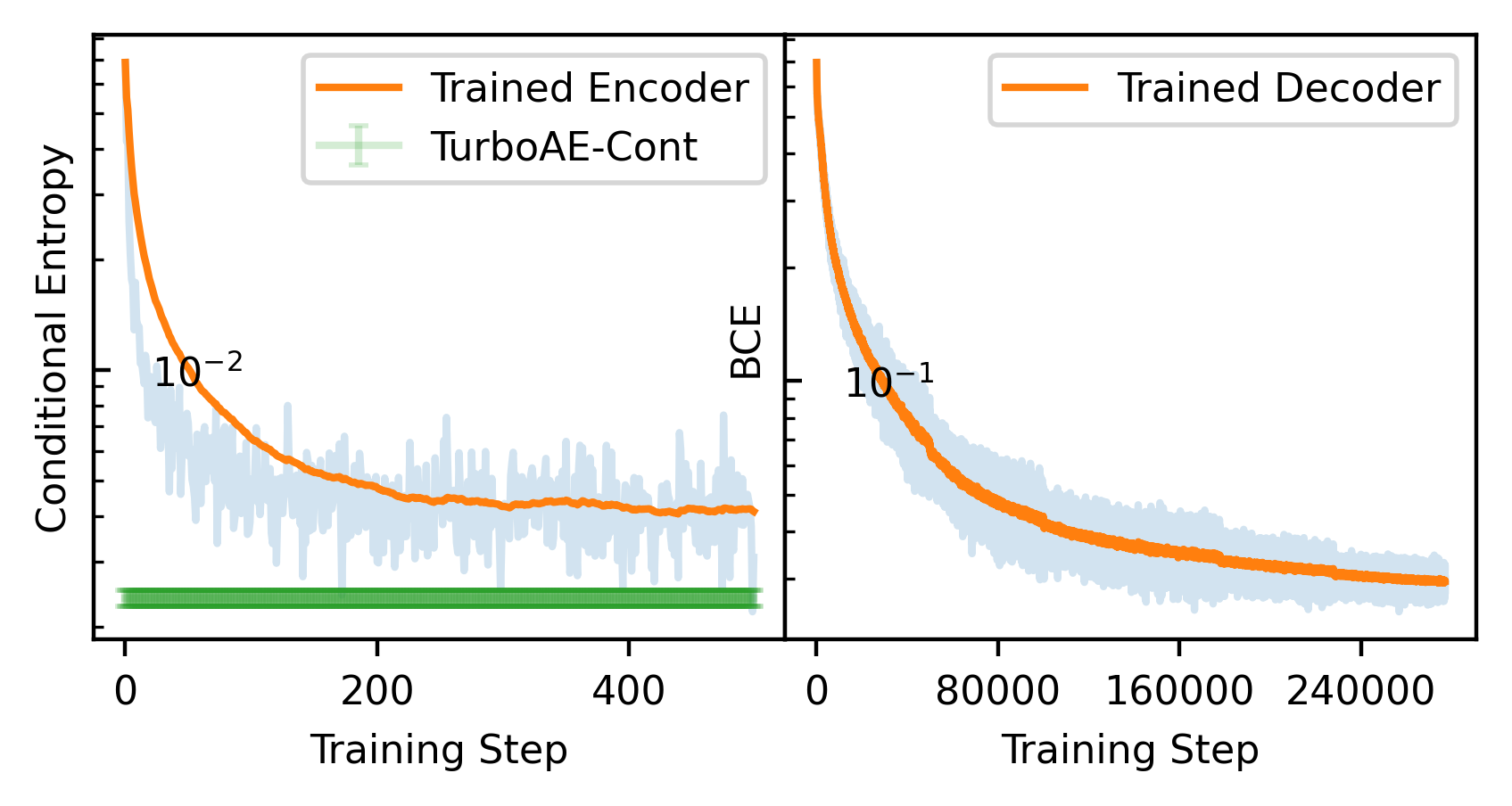}
    \caption{\small \textit{Left}:
    Learning curve of a Turbo-like encoder during training with conditional entropy at block length 16. The green bar shows the estimated conditional entropy of TurboAE-cont at the same block length. \textit{Right}: Learning curve of a neural decoder paired with the learned encoder from the left panel. Training was done at block length 100 and used binary cross-entropy as the cost function. The encoder was fixed during the training of the decoder.}
    \label{fig:enc_dec_training_curves}
\end{figure}

 See Fig.~\ref{fig:enc_dec_training_curves} as referred to in Section~\ref{section:training_encoders}.

\end{document}